\documentclass[showpacs,,twocolumn,superscriptaddress]{revtex4}

\usepackage{graphicx}
\usepackage{amsmath}
\usepackage{amssymb}
\usepackage{amsfonts}
\usepackage{bm}
\usepackage{color}

\def\bea{\begin{eqnarray}}
\def\eea{\end{eqnarray}}
\def\be{\begin{equation}}
\def\ee{\end{equation}}
\def\bes{\begin{subequations}}
\def\ees{\end{subequations}}
\def\nonum{\nonumber\\}
\renewcommand\Im{\operatorname{Im}}

\begin{document}

\title{\hspace{-1mm}{Quantum Coherence and Population Transfer in a Driven Cascade Three-Level Artificial Atom}}

\author{Sung Un Cho}
\altaffiliation{Present address: Department of Physics and Astronomy, Seoul National University, Seoul 151-747, Korea}
\affiliation{Korea Research Institute of Standards and Science, Daejeon 305-340, Korea}

\author{Han Seb Moon}
\affiliation{Department of Physics, Pusan National University, Busan 609-735, Korea}

\author{Young-Tak Chough}
\affiliation{Department of Healthcare $\&$ Medical Technology, Gwangju University,  Gwangju 503-703, Korea}

\author{Myung-Ho Bae}
\affiliation{Korea Research Institute of Standards and Science, Daejeon 305-340, Korea}

\author{Nam Kim}
\affiliation{Korea Research Institute of Standards and Science, Daejeon 305-340, Korea}

\date{\today}

\begin{abstract}
We present an experimental investigation on the spectral characteristics of an artificial atom ``transmon qubit'' constituting a three-level cascade system ($\Xi$-system) in the presence of a pair of external driving fields. We observe two different types of Autler-Townes (AT) splitting: type I, where the phenomenon of two-photon resonance tends to diminish as the coupling field strength increases, and type II, where this phenomenon mostly stays constant. We find that the types are determined by the cooperative effect of the decay rates and the field strengths. Theoretically analyzing the density-matrix elements in the weak-field limit where the AT effect is suppressed, we single out events of pure two-photon coherence occurring owing to constructive quantum interference.
 \end{abstract}

\maketitle

\section{Introduction}
Recently,  circuit-quantum electrodynamics (c-QED) three-level systems have attracted much attention 
\cite{AT_wallraff,mika_AT,kelly, EIT_tsai,3levela, delsing_prl,nakamura_2013,3levelb,3levelc,AT_mary}\, because they are ideal solid-state systems to investigate quantum interference phenomena, for instance,  electromagnetically induced transparency (EIT) \cite{EIT,Harris}.
The EIT phenomenon is a key factor in such a three-level system mainly because of its potential application in quantum information processing as a photonic information storage device \cite{RMP_2010,lovovsky}. 
However, when field strengths are large, the EIT can be mixed with, or affected by, the so-called Autler-Townes (AT) effect \cite{AT,Cohen}, which is a generic type of ac-Stark splitting. 
For this reason, the experimental discrimination between these two different yet similar effects becomes an important task \cite{sanders_prl}. 
In fact, recent studies \cite{moon_opt, moon_pra} suggest possible methods to discriminate  between multi-photon processes and to single out the pure two-photon process, which is a result of constructive or destructive quantum interference, depending on the ratio of the decay rates \cite{Berman, agarwal}. 
Nevertheless experimental analyses on the quantum interference phenomena in terms of decay rates for c-QED systems have seldom been carried out.   

\begin{figure} [!b]
\includegraphics[width=3.2in]{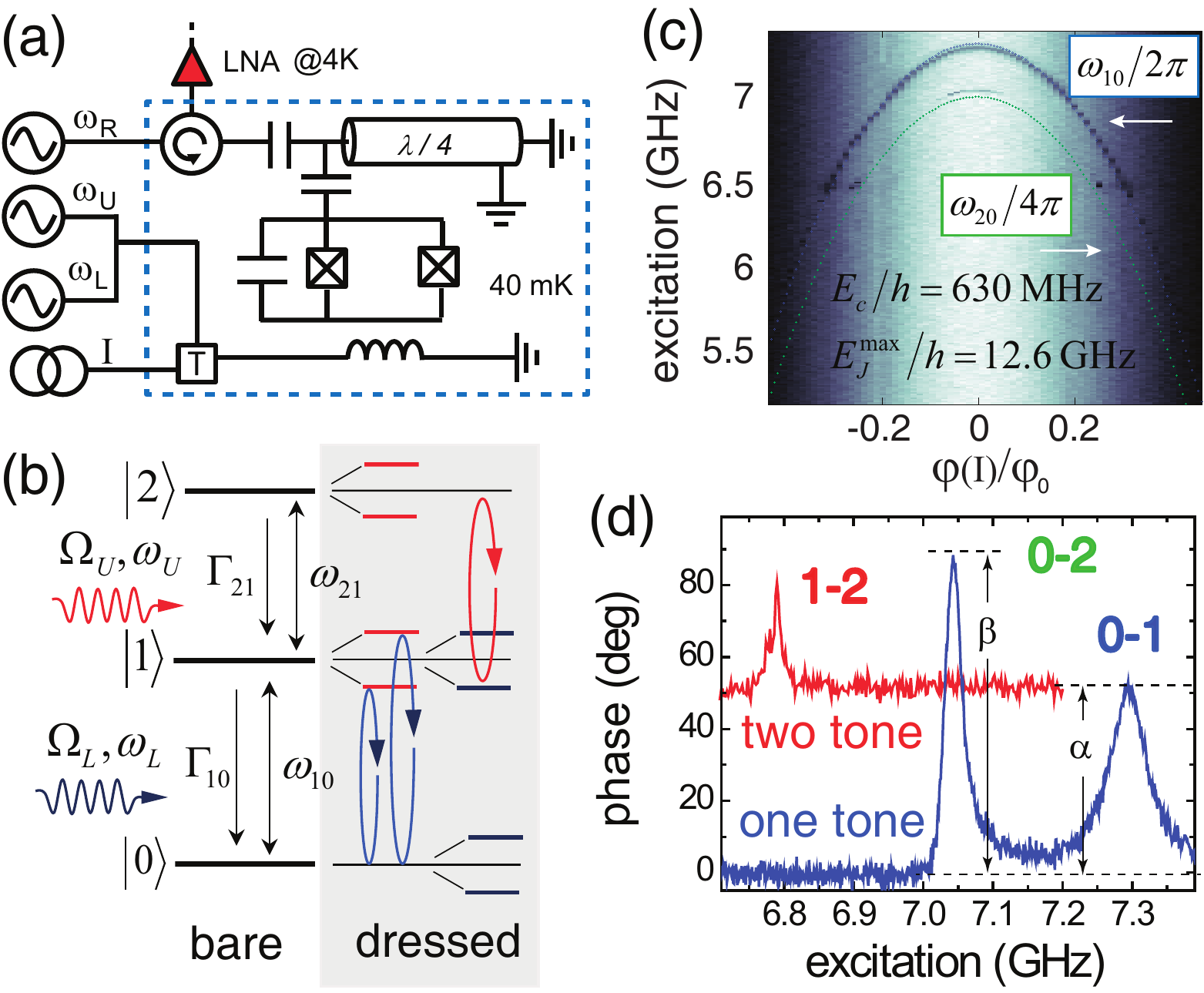}
\caption{(Color online) (a) Schematic diagram of the measurement set-up. Each level transition frequency is modulated by flux bias $I$ and the qubit is excited with excitation frequencies $\omega_L$ and $\omega_U$, which are combined at room temperature with 10 dB attenuators and then further attenuated by 60 dB at cold stages (70 dB in sum). (b) Energy diagram of a $\Xi$-system in bare states, which are dressed by coupling fields  ${\Omega}_U$ and ${\Omega}_L$. (c) Cavity responses to the applied frequency resonant with the transition frequencies  $\omega_{10}$(blue dots) and $\omega_{20}/2$(green dots), in agreement with the theory. (d) The excitation frequency profile at flux $\varphi$ = 0. Single-tone spectrum (blue) with strong field and two-tone spectrum (red). The two-tone spectrum is obtained by frequency sweep while applying a continuous wave tuned to the transition frequency $\omega_{10}$.}
\label{fig1}
\end{figure} 

In our work, we experimentally investigate the characteristics of the spectral splitting phenomenon occurring in a three-level cascade system ($\Xi$-system) driven by a pair of external fields. The three-level artificial atom is realized by a transmon qubit  which is integrated in a superconducting cavity \cite{koch2007}. We find that there are two generically different types of AT splitting cooperatively depending on the relative strength of the two driving fields and the relative decay rates. In the weak-field limit where the AT effect is suppressed, we analyze the measured level population by decomposing the density matrix elements and find that a constructive, rather than destructive, interference phenomenon occurs in the configuration of the decay rates in our system. 
Considering that the quantum interference phenomena are rarely observed in a single-atom level \cite{single atom}, and are mostly seen in ensemble states of a many-atom system in quantum optics \cite{EIT,ensemble}, our work is very significant in itself although it is done in an artificial single-atom system.
Furthermore our work has also shown that an appropriate pumping-probe configuration with the right decay-rate ratio is required to establish an EIT device for future application in c-QED systems.

\section{Experimental setup}

Figure 1(a) depicts our experimental setup.
The sample is fabricated on a 300-nm-thick SiO$_2$ layer thermally grown on a high-purity silicon substrate through conventional electron-beam lithography and aluminum double-angle evaporation. 
The qubit is located on the voltage antinode of a quarter-wavelength coplanar waveguide resonator \cite{na_trans} with a resonance frequency of $\omega_R/2\pi\approx5.05$ GHz and an energy damping rate of $\Gamma_R/2\pi\approx4$ MHz. 
Figure 1(b) specifies various symbols for our $\Xi$-system, 
defining the relaxation rates $\Gamma_{10}$ and $\Gamma_{21}$ from levels $|1\rangle$ and $|2\rangle$, respectively, 
as well as the energy gaps, $\hbar\omega_{jk}= E_j-E_k$, etc. 
The system is maintained at a temperature of 40 mK on the mixing chamber of a dilution refrigerator. 

The vacuum Rabi splitting for the lower-level transition is $g_1/2\pi\approx100$ MHz, 
whereas the coupling rate of the upper level transition is $g_2/2\pi\approx140$ MHz from spectroscopy, 
which approximates to the transmon-resonator coupling strength ratio $g_{j+1}/g_1\sim\sqrt{j+1}$.

We record phase changes by the dispersive shift of resonance frequency \cite{AT_wallraff} from the reflection signal ($S_{11}$) with a vector network analyzer. The average photon numbers of cavity probe tone are less than unity to obtain the spectra of  $\omega_{10}$ and $\omega_{20}$ over the flux bias, as shown in Fig.\ 1(c). 

We extract the charge energy $E_C/h\approx 0.63$ GHz and the maximum Josephson energy $E^\textup{max}_J/h\approx 12.6$\, GHz based on a model for transmon qubit \cite{koch2007}. 
We tune $E_J/ E_C$ to be its maximum at $\varphi/\varphi_0$ = 0 or 1 ($\varphi_0$ is the magnetic-flux quantum) to  reduce the cavity effect  which can modify the spontaneous decay rates, i.e., the Purcell effect \cite{purcell}, meaning that we have  $\omega_{10}/2\pi=7.3$ GHz and $\omega_{21}/2\pi=6.8$ GHz. 
The two-photon process for $\omega_{20}=\omega_{10}+\omega_{21}$ can be identified from the separate measurements of single-tone (blue line) and two-tone  (red line) spectroscopy in the saturation limit of a dispersive shift in the cavity frequency, as shown in Fig.\ 1(d). 
Here, we also obtain the heights of the peaks $\alpha$ and $\beta$ in the saturation limits, which  will determine the ratio of the level populations via $S_{11} \sim A\rho_{11} + B\rho_{22}$, where $\rho_{ii}$ is the density-matrix element for energy level $i$, and $A$ and $B$ are weighting factors (this is discussed later).

On the other hand, we obtain the decay rates of each level from time domain measurements based on the theory of damped Rabi oscillations \cite{kosugi_prb,kosugi_con}. 
The results indicate that $\Gamma_{10}/2\pi\approx0.44$ MHz,
$\gamma_{10}^\varphi/2\pi\approx0.99$ MHz, $\Gamma_{21}/2\pi\approx1.63$ MHz, and $\gamma_{20}^\varphi/2\pi\approx1.89$ MHz, where $\Gamma_{ij}$ and $\gamma_{ij}^\varphi$ are the relaxation rate and the pure dephasing rate from states $|i\rangle$  to $|j\rangle$, respectively \cite{rabi}. 

\begin{figure}[b]
\centering     
\includegraphics[width=3.2in]{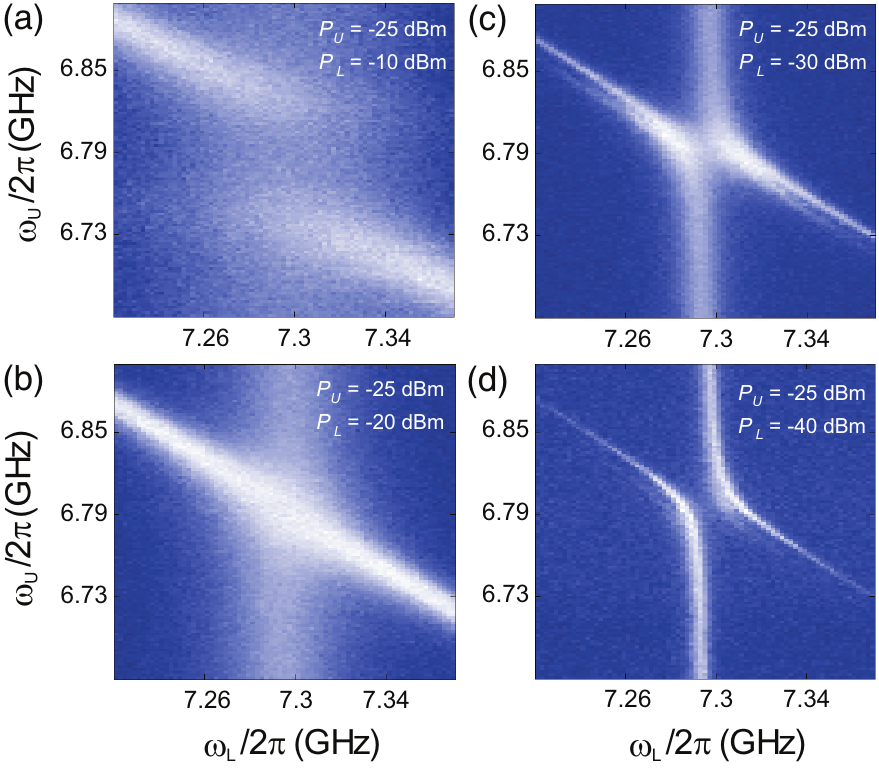}
\caption{(Color online) The AT avoided crossing for a few combinations of driving power $P_{L(U)}$ (source output reading) as a function of $\omega_L$ and $\omega_U$. Rabi frequencies corresponding to the respective field strength are $\Omega_L/2\pi$= (a) 95.4, (b) 30.2, (c) 9.5, and (d) 3.0 MHz for a fixed upper field strength of $\Omega_U/2\pi$= 26.5 MHz.} 
\label{fig2}
\end{figure}  
\begin{figure*}
\includegraphics[width=2\columnwidth]{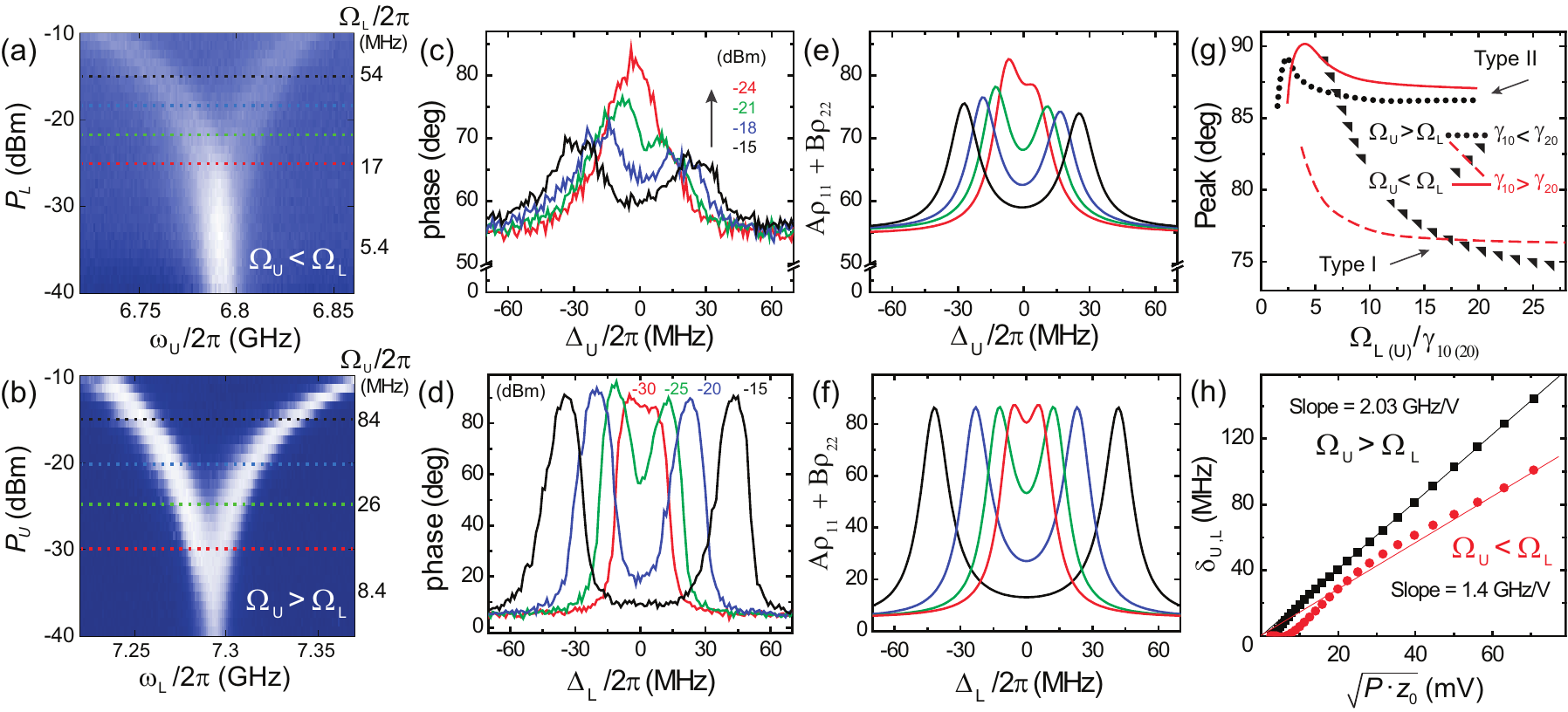}
\caption{(Color)  (a),(b) Density plots of phase change caused by resonance shift over the pump power $P_{L(U)}$ 
(instrument output reading value) 
and probe frequency $\omega_{U(L)}$, resulting in the AT splitting. 
(c),(d) Spectra of the phase changes for various driving powers corresponding to the horizontal dotted lines in (a) and (b), respectively, over detuning  $\Delta_L=\omega_{10}-\omega_L$ and $\Delta_U=\omega_{21}-\omega_U$. 
(e) and (f) are the simulation results corresponding to (c) and (d), respectively. In (e), the pump frequency $\omega_L$ is blue detuned by 2$\pi\times$2 MHz from transition frequency $\omega_{10}$ to fit the asymmetries shown in (c). 
(g) Variation of the peak heights in (e) and (f) vs the ratio of the pump strength to the decay rate of the pumped transition pair (black solid circles and triangles). Solid and dashed red lines represent the inverted ratio $\gamma_{20}/\gamma_{10}$. 
(h) The peak separation ($\delta_{U,L}$) as a function of the pump power $P$ for two driving field configurations, with characteristic impedance $z_0$ = 50\,$\Omega$.}
\label{fig3}
\end{figure*}

\section{Avoided\, Crossing}

After the experiment is set up as described, we carry out the level population mapping over the ($\omega_U,\omega_L$) plane for varying values of $\Omega_L$ while keeping the value of $\Omega_U$ fixed, as shown in Figs.\ 2(a)--2(d). 
The archetypal quantum behavior of avoided crossing \cite{tak_2000,Bishop2008}\, is manifested throughout the figures, which is, of course, a direct result of the level splitting induced by the driving fields. However, a much more interesting feature is that the direction of the anticrossing borderline turns by 90 degrees, as seen from Figs.\ 2(a)--2(d). 
This rotation of lines implies that the AT splitting is brought about by the stronger field of the two because the stronger field acts as ``pump,'' whereas the weaker field acts as ``probe.'' Therefore, if  $\Omega_L$ is greater than $\Omega_U$, the former splits level $|1\rangle$  and the latter probes this splitting, and vice versa. 
In our view, this interesting feature has not been clearly pointed out in the literature. 
When the two Rabi frequencies are comparable to each other, i.e., $\Omega_U\approx\Omega_L$  as shown in Fig.\ 2(b), level $|1\rangle$\,  suffers double splitting induced by the two fields (called supersplitting by some) \cite{Bishop2008}, which eventually blurs the spectrum around the bare resonance point. 
The corresponding numerical-calculation results, based on the master-equation formalism \cite{dam1992,tian1992,Carmichael1993,jian_prb}, 
reproduce our results very well 
(refer to Appendix A for further discussions on the numerical methods and simulation results).

\section{Characterization\, of\, AT\, splitting}
We now compare the responses of the system when (1) the lower transition is pumped and (2) the upper transition is pumped, as shown in Figs.\ 3(a) and 3(b), respectively. To obtain the responses presented in Fig.\ 3(a), we fix $\Omega_U/2\pi=8.5$ MHz and scan the probe frequency $\omega_U$ around upper transition $\omega_{21}$ for increasing values of $\Omega_L$, such that $\Omega_L>\Omega_U$, with the pump frequency tuned to the lower transition ($\omega_L=\omega_{10}$). 
To obtain the responses presented in Fig.\ 3(b), we fix $\Omega_L/2\pi=5.3$ MHz and scan the probe frequency $\omega_L$ around lower transition $\omega_{10}$
for increasing values of $\Omega_U$, such that $\Omega_U>\Omega_L$, with the pump frequency tuned to the upper transition ($\omega_U=\omega_{21}$). In both cases, as the pump strength [i.e., $\Omega_L$ in Fig.\ 3(a) and $\Omega_U$ in Fig.\ 3(b)] increases, the spectral splitting widens as expected. However, we can clearly see a striking difference between the two cases in that the brightness of the peaks diminishes 
in Fig.\ 3(a), whereas it remains essentially constant in Fig.\ 3(b). 
Let us therefore call the case of Fig.\ 3(a) type I and the case of Fig.\ 3(b) type II. 
This feature is quantitatively demonstrated in Figs.\ 3(c) and 3(d). The colors of each of the lines 
in Figs.\ 3(c) and 3(d) correspond to those of the horizontal dots indicating the pump strength in Figs.\ 3(a) and 3(b), respectively. 
Figures 3(e) and 3(f) present the numerical calculations for quantity $A\rho_{11}+B\rho_{22}$ for both cases,
agreeing well with the experimental data presented in  Figs.\ 3(c) and 3(d), respectively. 
The weighting factors $A$ = 1.06 and $B$ = 1.65 are extracted from the resonance pull of the cavity using the relationship $\alpha/\beta=A/B$ \cite{tomography_wallraff}, where $\alpha$ and $\beta$ are defined in Fig.\ 1(d). 

Then, we investigate the cause for the significant difference between type I and type II. 
We first assumed that this difference may occur due to the corresponding pump configuration. 
However, further investigation indicates that the atomic decay rates play an important role as well.
First, we define the decay rates, $\gamma_{10}/2\pi\equiv(\Gamma_{10}+\gamma^{\varphi}_{10})/2\pi\approx1.43$ MHz 
and $\gamma_{20}/2\pi\equiv(\Gamma_{21}+\gamma^{\varphi}_{20})/2\pi\approx3.52$ MHz.
Therefore, obviously  $\gamma_{20}>\gamma_{10}$ in our system. 
Given this case, we trace the heights of the resonance peaks under the conditions presented in Figs.\ 3(e) and 3(f) by scanning the corresponding pump strengths. 
The result is indicated by the solid circles and triangles in Fig.\ 3(g). 
The solid triangles in Fig.\ 3(g) correspond to the condition  in Fig.\ 3(e), i.e.,\, $\Omega_U<\Omega_L$, which shows the peak heights decreasing with increasing $\Omega_L$ values, agreeing well with the trend shown in Fig.\ 3(e), i.e., type I. On the other hand, the solid circles corresponding to the condition in Fig. 3(f), i.e., $\Omega_U>\Omega_L$, show the peak heights hardly varying over a wide range of $\Omega_U$ values, agreeing well with the trend in Fig.\ 3(f), i.e., type II. The red solid and dotted lines in Fig.\ 3(g) are obtained by just reversing the size of the decay rates, i.e.,\,$\gamma_{20}<\gamma_{10}$. 
The red solid line corresponds to the condition $\Omega_U<\Omega_L$, and the red dashed line corresponds to $\Omega_U>\Omega_L$. 
Quite surprisingly, the result indicates that the types of splitting are essentially interchanged upon reversal of the corresponding decay strengths. 
The underlying physics of this interesting phenomenon must be simple but it is yet to be explored theoretically. 
In Fig.\ 3(h), we measure the dependence of the peak separations $\delta_{L,U}$  on field power $P$ from Figs.\ 3(a) and 3(b). 
The results for both cases are linear as expected in the strong field regime because $\Omega_{U(L)}=\delta_{U(L)}$, giving a slope ratio of $\sim$1.4 (= 2.0\,GHz$\cdot$V$^{-1}$/1.4\,GHz$\cdot$V$^{-1}$), which is another manifestation of the ratio $g_2/g_1\sim\sqrt{2}$. 
Thus, the field power unit is translated to Rabi frequency via linear extrapolation to amplitude zero. This agrees well with the model introduced in Ref. \cite{AT_wallraff}.

\section{Quantum\, interference}

\begin{figure}
\includegraphics[width=2.4in]{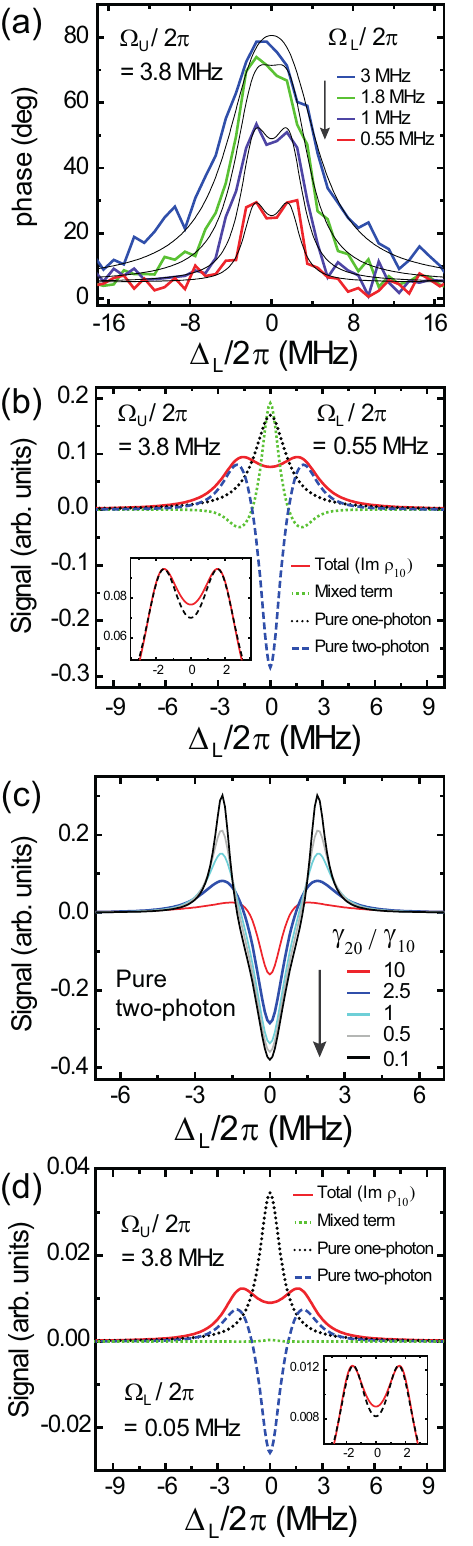}
\caption{(Color) (a) The phase change caused by the resonance shift for various probe Rabi frequencies $\Omega_L$. (b) Decomposition of the absorption spectrum Im$\rho_{10}$ into three terms as mentioned in the text, for the experimental parameters $\Omega_U/2\pi$ = 3.8 MHz and $\Omega_L/2\pi$ = 0.55 MHz. (c) Spectral behavior of pure two-photon process for various relative decay rates. The blue line corresponds to our experimental condition of $\gamma_{20}/\gamma_{10}$\, = \,2.5. (d) Changes in the three lines in (b) for a much weaker value of $\Omega_L/2\pi$= 0.05 MHz. Insets in (b) and (d) are double Lorentzian fits (black dashed line) for Im$\rho_{10}$.}
\label{fig4}
\end{figure}

In order to observe the sheer quantum coherence effects, 
we investigate the weak-field regime, where the Rabi frequencies are not greater than the decay rates of the system, so as to avoid complications such as the power splitting discussed previously. 
To obtain the results presented in Fig.\ 4(a), we fix the pump strength $\Omega_U/2\pi$ = 3.8 MHz at resonant with the upper transition frequency, i.e., $\omega_U=\omega_{21}$, given the total decay rate of the system, $\gamma_{21}/2\pi\equiv(\gamma_{10}+\gamma_{20})/2\pi = 4.95$ MHz.
Then, we scan the probe frequency $\omega_L$ around the lower transition $\omega_{10}$
for varying values of probe strength $\Omega_L$, such that $\Omega_L<\Omega_U<\gamma_{21}$.
The thin smooth lines are theoretical curves corresponding to  $A\rho_{11}+B\rho_{22}$ with the aforementioned values of $A$ and $B$. 
A dip clearly develops as the probe strength decreases, such that $\Omega_L/\gamma_{10}<1$.
We discuss the cause for the formation of this dip in the following paragraphs. 

The population profile of the probe ($\sim S_{11}$) is known to be proportional to Im$\rho_{10}$ \cite{scully97}. 
We decompose the corresponding theoretical spectrum of Im$\rho_{10}$ for the red line in Fig.\ 4(a), i.e., the line for $\Omega_L/2\pi$ = 0.55 MHz, 
into three components---pure one-photon coherence, pure two-photon coherence, and mixed coherence component---as shown 
in Fig.\ 4(b); this is a strategy developed in the previous studies on three-level systems \cite{moon_opt, moon_pra}, further details of which are presented in Appendix B.
The black dotted line represents one-photon coherence, i.e., the step-by-step excitation from the ground state to the level top
via the intermediate level. On the other hand, the pure two-photon coherence represented by the blue dashed line is the process in which the level populations do not change. The mixed coherence component represented by the green dotted line is determined by subtracting the values of the one-photon and pure two-photon coherences from the total absorption strength. 
This is relevant to the two-photon absorption process where the level population oscillates between the ground state and the level-top.
Figure 4(b) clearly shows that the pure two-photon coherence process (blue dashed line) causes the dip at the center.

We now investigate the variation of the pure two-photon coherence signal against the relative decay rate $\gamma_{20}/\gamma_{10}$, 
as shown in Fig.\ 4(c), where we fix $\gamma_{10}$ at the value of our qubit, and vary $\gamma_{20}$. 
The blue line ($\gamma_{20}/\gamma_{10}=2.5$) in the figure corresponds to the experimental situation of our qubit. 

\begin{figure*}
\includegraphics[width=6.4in]{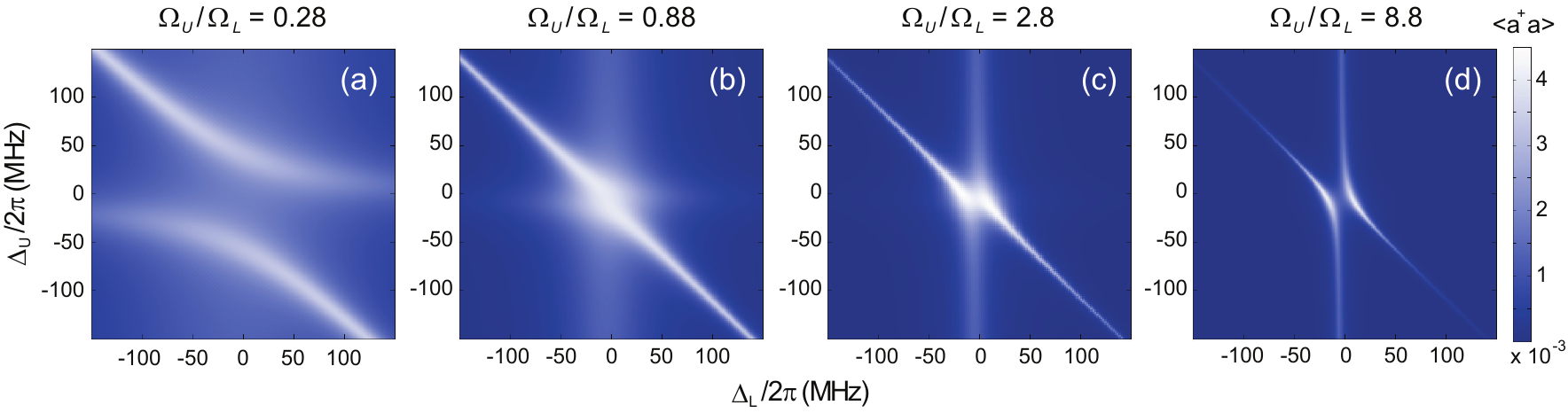}
\caption{(Color online) Numerically calculated behavior of the mean intracavity photon number $\langle a^\dagger a\rangle$ on the ($\omega_L,\omega_U$) plane centered at ($\omega_{10},\omega_{21}$). Parameters in the calculation in each plot correspond to those of Figs.\ 2(a)--2(d), in the main text.}
\label{fig 5}
\end{figure*}
The figure clearly shows that the dip gets deeper as the ratio $\gamma_{20}/\gamma_{10}$ decreases. 
It is known that destructive (constructive) interference is expected when the decay ratio $\gamma_{20}/\gamma_{10}$ is smaller (larger) than unity \cite{Berman, agarwal}. 
Therefore, according to the theory, for the condition, $\gamma_{20}/\gamma_{10} = 2.5$, constructive interference is expected to occur in our system.
Even though the calculated two-photon coherence terms have not shown a qualitative change, 
they provide reasonable evidence for the constructive interference.
The inset in Fig.\ 4(b) indicates that the depth of the total spectrum dip for our experimental condition, 
i.e., for $\gamma_{20}/\gamma_{10} = 2.5$, is shallower than the double Lorentzian dip (black dashed line)
which is another signature of the constructive interference, 
although the difference is small where we assume double Lorentzian fits the AT splitting.

Considering the possibility that the non-negligible probe strength may wash out the coherence and result in poor resolution, we look at the case of an even smaller value of probe strength $\Omega_L/2\pi = 0.05$ MHz, as shown in Fig.\ 4(d).
There are no significant differences observed in comparison to Fig.\ 4(b) 
except for the mixed term which is influenced mainly by the probe field intensity.
Thus, we conclude that the experimentally observed spectral dip in Fig.\ 4(a) is most likely due to constructive quantum interference.
We have come to the conclusion of constructive interference 
taking into account both spectra of absorption and pure two photon altogether, as discussed above.

On the other hand, when the decay-rate ratio is reverted as $\gamma_{20}/\gamma_{10} = 0.1$, for instance,
a Fano-profile dip due to the destructive interference is theoretically predicted in Appendix C.  
Currently, quantum interference in the case of lower-level driving is yet to be explored because the spectral signal still remains unresolved 
in the weak-field limit. 
In this configuration, only constructive interference is theoretically expected to occur, regardless of the ratio of the decay rates \cite{Berman, agarwal}.

\section{Conclusion}

We investigated the Autler-Townes splitting effects in the strong-field limit in a three-level cascade system of an artificial atom. We found that both the relative strength between the applied fields and the relative level decay rates cooperatively determine the characteristics of the Autler-Townes spectrum. In the weak-field limit, we analyzed the resonance dips by decomposing the density matrix into three different coherence components, thereby discriminating between the multiphoton processes. We have singled out the event of a pure two-photon coherence process and observed constructive quantum interference in the atomic spectrum.

\begin{acknowledgments}
We thank P. D. Nation, M.-J. Hwang, M.-S. Choi, and K. C. Kang for valuable discussion. This work was partially supported  by the Korea Research Institute of Standards and Science under the project ``Convergence Science and Technology for Measurements at the Nanoscale,'' Grant No. 13011041 and the National Research Foundation of Korea (NRF) with funding from the Ministry of Education (Grants No. 2012R1A2A1A01006579 and  No. 12A12168451).
\end{acknowledgments}

\appendix

\section{Survey on cavity effects} 

The Hamiltonian for the qubit of a three-level ($|0\rangle,|1\rangle,|2\rangle$)\,$\Xi$-system, pumped by a pair of external fields (we denote them ${\cal E}_U$ and ${\cal E}_L$), interacting with a cavity mode is given by
\bea
H &=& \hbar\omega_C   a^\dagger   a + \hbar\omega_{10}|1\rangle\langle1|+\hbar(\omega_{10}+\omega_{21})|2\rangle\langle2|\nonum
&&+i\hbar g_1 \left(a^\dagger|0\rangle\langle1| - a|1\rangle\langle0|\right)+i\hbar g_2 \left(  a^\dagger|1\rangle\langle2|-  a|2\rangle\langle1|\right)\nonum
&&+i\hbar\left(\Omega_L\over2\right) \left(e^{-i\omega_Lt}|1\rangle\langle0|-e^{i\omega_Lt}|0\rangle\langle1|\right)\nonum
&&+i\hbar\left(\Omega_U\over2\right) \left(e^{-i\omega_Ut}|2\rangle\langle1|-e^{i\omega_Ut}|1\rangle\langle2|\right)
\eea\normalsize
under DA (dipole approximation) and RWA (rotating-wave approximation), where $\omega_C$ is the cavity resonant frequency; $a^\dagger  (a)$ is the boson creation (annihilation) operator for the cavity mode;
$\omega_{j,k}$ is the transition frequency between the qubit energy levels $|j\rangle$ and $|k\rangle$; $g_j$ is the dipole coupling strength between the resonator mode and the level transition between $|j\rangle$ and $|j-1\rangle$; and $\Omega_{U(L)}$ and $\omega_{U(L)}$ are the Rabi frequency and the angular frequency of the external field ${\cal E}_{U(L)}$, respectively.

Without a cavity driving field, the response of the system is given by the steady-state solution of the Markovian master-equation formalism \cite{tian1992, dam1992, Carmichael1993, jian_prb}, i.e.,
\bea
\dot\rho&=&-{1\over\hbar}\left[H, \rho\right]\nonum
&&+\sum^2_{i=1}{\Gamma_{i,i-1}\over2}\left(2|i-1\rangle\langle i|\rho|i\rangle\langle i-1|-|i\rangle\langle i|\rho-\rho|i\rangle\langle i|\right)\nonum
&&+\sum^2_{i,j=0,(i\ne j)}{\gamma_{ij}^\varphi\over2}|i\rangle\langle i|\rho|j\rangle\langle j|\nonum
&&+{\Gamma_C\over2}\left(2a\rho a^\dagger -a^\dagger a\rho-\rho a^\dagger a\right)
\eea
where $\Gamma_C$ is the cavity photon decay, $\Gamma_{i,i-1}$ is the relaxation rate from state $|i\rangle$ to $|i-1\rangle$, and $\gamma_{i,j}^\varphi$ is the pure dephasing rate.

Care must be taken in solving the master equation, given by Eq.\ (A2), in order not to accumulate the numerical error in view of the fact that the system is involved with so many different frequencies, which are highly detuned from each other---particularly when the cavity is turned on.
The complexity of the problem is quadrupled, when cavity interaction is involved, and accordingly the computing time increases markedly. We resort to the International Mathematics and Statistics Library (IMSL) subroutine DIVPAG, one of the double precision ordinary differential equation (ODE) solvers of the initial-value problems. 
Figures 5(a)--5(d) present the behavior of the mean intracavity photon number $\langle a^\dagger a\rangle$ on the ($\Delta_U,\Delta_L$) plane, where $\Delta_U\equiv\omega_{21}-\omega_U$  and $\Delta_L\equiv\omega_{10}-\omega_L$.
Although the exchanges of excitation between the qubit and cavity mode are highly unlikely in the dispersive coupling regime, there is still nonzero probability of excitation of the cavity field.
Furthermore, the spectral behavior of $\langle a^\dagger a\rangle$  is essentially identical to that of the qubit excitation, in comparison to Figs.\ 2(a)--2(d) in our main text, except for the size of the excitation which is merely of the order of $10^{-3}$ photons.  

However, the numerical simulations for the system with and without the cavity indicate that the role of cavity 
is negligible owing to its extreme far-off resonance from the atomic transitions, 
other than slightly shifting the spectral lines as a whole from that of the system 
without the cavity by an amount precisely predicted by the dispersive Hamiltonian \cite{blais_2004, koch2007}. 
In the dispersive coupling regime, the level shift owing to the cavity effect is given 
by $\chi_j = g^2_j/(\omega_{j,j-1}-\omega_C)$ \cite{koch2007, tak_2000}. 
For convenience, we treat the dressed frequency ${\omega'}_{j,j-1}\equiv\omega_{j,j-1}+\chi_j$  
as the ``bare'' transition frequency $\omega_{j,j-1}$.

\section{Decomposition of the three components of quantum coherence} 

For a ladder-type three-level configuration, the density-matrix equations are given by

\bes
\begin{align}
\dot\rho_{00} &= \gamma_{10}\rho_{11}+\Omega_L \Im \rho_{10} \\
\dot\rho_{11} &= \gamma_{20}\rho_{22}+\Omega_U \Im \rho_{21}-\Omega_L \Im \rho_{10} \\
\dot\rho_{22} &= -\gamma_{20}\rho_{22}-\Omega_U \Im \rho_{21} \\
\dot\rho_{10} &= \left(i/ 2\right)\left[\Delta_1\rho_{10}-\Omega_U \rho_{20}+\Omega_L\left( \rho_{11}-\rho_{00}\right)\right] \\
\dot\rho_{20} &= \left(i/ 2\right)\left[\Delta_0\rho_{20}+\Omega_L \rho_{21}-\Omega_U\rho_{10}\right] \\
\dot\rho_{21} &= \left(i/ 2\right)\left[\Delta_2\rho_{21}+\Omega_L \rho_{20}+\Omega_U\left( \rho_{22}-\rho_{11}\right)\right] 
\end{align}
\ees\normalsize
where $\Delta_0 = 2(\Delta_L+\Delta_U)+i\gamma_{20}$, 
$\Delta_1=2\Delta_L+i\gamma_{10}$, 
and $\Delta_2=2\Delta_U+i(\gamma_{10}+\gamma_{20})$, 
and $\Delta_L$ and $\Delta_U$ are the detuning of the probe and pump fields from their own target levels, respectively. 

The decay rates \,$\gamma_{ij}=\gamma_{ji}$\, are the sum of the relaxation rate $\Gamma_{ij}$ and the pure dephasing rate \,$\gamma_{i0}^\varphi$. We neglect the relaxation rates induced by thermal excitations, i.e., \,$\Gamma_{01}=\Gamma_{12}=0$ and the direct relaxation rate from $|2\rangle$ to $|0\rangle$ in a three-level $\Xi$-system, i.e., \,$\Gamma_{20}=0$ .

Following the method introduced in our previous contributions \cite{moon_opt, moon_pra}, 
the one-photon resonance component is obtained by equating $\rho_{20}$ to zero in the above equations, such that
\be
\left(\Im\rho_{10}\right)_{\rm one}=\Im\left[{\Omega_L\over\Delta_1}\left(\rho^{(0)}_{00}-\rho^{(0)}_{11}\right)\right]
\ee\normalsize
where $\rho^{(0)}_{00}$ and $\rho^{(0)}_{11}$   
represent the populations of the $|0\rangle$ and $|1\rangle$ states under the condition $\rho_{20}=$\,0, respectively.
The pure two-photon coherence component is obtained when all of the populations of the intermediate and excited states are ignored, i.e., $\rho_{00}=$\,1 and $\rho_{11}=\rho_{22}=$\,0 such that

\be
\left(\Im\rho_{10}\right)_{\rm two}=\Im\left[{\Omega_L\over\Delta_1}\rho_{20}\right]
\ee\normalsize

The mixed coherence term is given by subtracting the one-photon and pure two-photon coherence components from {\rm Im}$\rho_{10}$, i.e.,
\be
\left(\Im\rho_{10}\right)_{\rm mixed}=\Im\rho_{10}-{\left(\Im\rho_{10}\right)}_{\rm one}-{\left(\Im\rho_{10}\right)}_{\rm two}
\ee\normalsize


\section{Quantum interference and relative decay rates}

Figure\ 6 shows two distinguished total absorption spectra; Fig. 6(a) shows a single merged peak while Fig. 6(b) shows an EIT-induced Fano-profile dip for the respective parameter values, which are written on the figures. 
The signal of the pure two-photon process shows an increase in amplitude as the relative decay rate changes
 in the way already mentioned in the main text.
Our simulation results are consistent with the author$'$s argument \cite{agarwal}
that  relative decay rates determine the characteristics of the quantum interference in a three-level $\Xi$-system, 
for instance, constructive interference for  $\gamma_{20}>\gamma_{10}$ or destructive interference  
for $\gamma_{20}<\gamma_{10}$
when the upper transition level is driven with the probe field applied to the lower one.
For the minimization of the probe field effects, we set $\Omega_L/2\pi$ to be as low as 0.05 MHz, while $\Omega_U/2\pi$ is set to be 1 MHz
to guarantee the two photon coherence effects. 
As simulation results illustrate, the condition of relative decay rate $\gamma_{20}>\gamma_{10}$ is required
for EIT to be observed in a three-level $\Xi$-system. 

\begin{figure}
\includegraphics[width=2.8in]{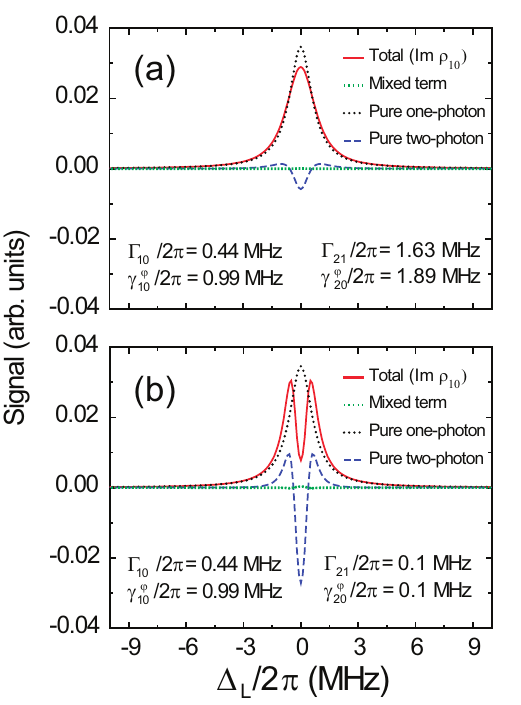}
\caption{(Color online)  Calculated absorption spectrum Im$\rho_{10}$ decomposed into three terms, as mentioned in the text, for two different relative decay rates (a) $\gamma_{20}/\gamma_{10}$ = \,1.9 and (b) $\gamma_{20}/\gamma_{10}$\, = \,0.1 for $\Omega_U/2\pi$ = 1.0 MHz and $\Omega_L/2\pi$ = 0.05 MHz.}
\label{fig 6}
\end{figure}


\begin{thebibliography}{99}
\bibitem{rabi}
In order to measure Rabi oscillation between levels $|2\rangle$ and $|1\rangle$
  , the lower level driving field is kept turned on to populate level
  $|1\rangle$.
\end{thebibliography}
\end{document}